\documentstyle[12pt]{article}   

\large
\newtheorem{definition}{Definition}[section]

\newcommand{\beq}{\begin{equation}}
\newcommand{\eeq}{\end{equation}}

\makeatletter
\@addtoreset{equation}{section}
\makeatother

\newtheorem{Theorem}{Theorem}[section]

\newtheorem{Lemma}{Lemma}[section]

\def\be{\begin{equation}}
\def\ee{\end{equation}}
\def\ba{\begin{eqnarray}}
\def\ea{\end{eqnarray}}

\def\agb{{\overline {{\cal A}/{\cal G}}}}

\def\Comp{{\mathchoice
{\setbox0=\hbox{$\displaystyle\rm C$}\hbox{\hbox to0pt
{\kern0.4\wd0\vrule height0.9\ht0\hss}\box0}}
{\setbox0=\hbox{$\textstyle\rm C$}\hbox{\hbox to0pt
{\kern0.4\wd0\vrule height0.9\ht0\hss}\box0}}
{\setbox0=\hbox{$\scriptstyle\rm C$}\hbox{\hbox to0pt
{\kern0.4\wd0\vrule height0.9\ht0\hss}\box0}}
{\setbox0=\hbox{$\scriptscriptstyle\rm C$}\hbox{\hbox to0pt
{\kern0.4\wd0\vrule height0.9\ht0\hss}\box0}}}}
\def\Co{{\mathchoice
{\setbox0=\hbox{$\displaystyle\rm C$}\hbox{\hbox to0pt
{\kern0.4\wd0\vrule height0.9\ht0\hss}\box0}}
{\setbox0=\hbox{$\textstyle\rm C$}\hbox{\hbox to0pt
{\kern0.4\wd0\vrule height0.9\ht0\hss}\box0}}
{\setbox0=\hbox{$\scriptstyle\rm C$}\hbox{\hbox to0pt
{\kern0.4\wd0\vrule height0.9\ht0\hss}\box0}}
{\setbox0=\hbox{$\scriptscriptstyle\rm C$}\hbox{\hbox to0pt
{\kern0.4\wd0\vrule height0.9\ht0\hss}\box0}}}}
\def\Rl{{\mathchoice
{\setbox0=\hbox{$\displaystyle\rm R$}\hbox{\hbox to0pt
{\kern0.4\wd0\vrule height0.9\ht0\hss}\box0}}
{\setbox0=\hbox{$\textstyle\rm R$}\hbox{\hbox to0pt
{\kern0.4\wd0\vrule height0.9\ht0\hss}\box0}}
{\setbox0=\hbox{$\scriptstyle\rm R$}\hbox{\hbox to0pt
{\kern0.4\wd0\vrule height0.9\ht0\hss}\box0}}
{\setbox0=\hbox{$\scriptscriptstyle\rm R$}\hbox{\hbox to0pt
{\kern0.4\wd0\vrule height0.9\ht0\hss}\box0}}}}

\title{The inverse loop transform}
\author{T. Thiemann\thanks{New address : Physics Department,
       Harvard University, Cambridge, MA 02138, USA, Internet :
       thiemann@abel.math.harvard.edu} \\
       Physics Department, The Pennsylvania State University,\\
       University Park, PA 16802-6300, USA}
\date{{\small Preprint CGPG-95/7-2, Preprint HUTMP-95/B-346}}

\begin{document}

\maketitle

\begin{abstract}
The loop transform in quantum gauge field theory can be recognized as the
Fourier transform (or characteristic functional) of a measure on the 
space of generalized 
connections modulo gauge transformations. Since this space is a compact 
Hausdorff space, conversely, we know from the Riesz-Markov theorem that
every positive linear functional on the space of continuous functions 
thereon qualifies 
as the loop transform of a regular Borel measure on the moduli space.\\
In the present article we show how one can compute the finite joint 
distributions of a given characteristic functional, that is, we derive
the inverse loop transform. 
\end{abstract}

\section{Introduction}

Recently, there has been made considerable progress in the development of a 
rigorous calculus on the space of generalized (that is, distributional)
connections modulo gauge transformations $\agb$ for quantum gauge field 
theories 
based on compact gauge groups \cite{1,2,3,4,5,6,7a,7b,8,9,10,11}. These 
developments can be summarized roughly as follows :\\
$\bullet$ The space of histories for quantum gauge field theory arises as 
the Gel'fand spectrum of the $C^*$ algebra generated by the Wilson loop 
functionals \cite{1}.\\
$\bullet$ There are two equivalent, useful descriptions of that spectrum :\\
a) it can be recognized as the set of {\em all} homomorphisms from the
group of loops into the gauge group \cite{2}.\\
b) it arises as the projective limit of the measurable spaces defined by
restricting the homomorphisms  to the cylindrical subspaces defined by 
piecewise analytical graphs $\gamma$ \cite{5,6}.\\
$\bullet$ The integral calculus on $\agb$ is governed by the fact that the
spectrum is always a compact Hausdorff space so that regular 
($\sigma$-additive) Borel measures thereon are in one to one correspondence
with positive linear functionals on $C(\agb)$ \cite{1}. Interestingly,
even diffeomorphism invariant measures can be constructed thereon \cite{2,3}.
Furthermore, the contact with constructive quantum gauge field theory is made
by the so-called loop transform which is nothing else than the characteristic
functional of the given measure \cite{9,10}, the role of the usual 
Bochner theorem \cite{13} being played by the Riesz-Markov theorem.\\
$\bullet$ Even differential geometry can be developed on cylindrical 
subspaces of $\agb$ \cite{7a,7b,8}.\\
$\bullet$ The motivation of the authors involved in these developments 
comes from quantum gravity formulated as a dynamical theory of connections
\cite{13a}.
The mathematical progress made has given rise to some new results \cite{11}
in quantum gravity. While this is a theory of complex-valued connections,
contact with the above mentioned results and techniques can be made by
a coherent state transform \cite{12}.\\
The correct version of that transform was found in \cite{12a} for the case
of pure gravity and extended in \cite{12b} to incorporate matter.\\
\\
In this paper we make yet another contribution to this subject, namely
we construct the analogue of the inverse Fourier transform on $\Rl^n$
\cite{14} which we call the ``inverse loop transform". This will allow
us to reconstruct a measure as defined by its finite joint distributions
from its characteristic functional. In particular, given a (singular)
knot invariant satisfying certain additional conditions, we can find
out to what measure it corresponds.\\
The plan of the paper is as follows :

In section 2 we recall the basic notions from calculus on $\agb$ as far
as necessary for the present context. The interested reader is referred 
to the article \cite{11} for further details.

In section 3 we first formulate and prove the analogue of the inverse 
Fourier transform theorem for compact gauge groups and then give a new 
definition of the loop transform based on the
notion of ``loop networks" \cite{15,16}. This notion will prove useful
in proving the inverse loop transform.

In section 4 we conclude by displaying the inverse loop transform of
physically interesting characteristic functionals.

In an appendix we indicate how all the results of this paper can be
rewritten in terms of edges rather than loops which is sometimes more
convenient in applications beyond integration on $\agb$. In particular
we introduce the notion of an ``edge-network" which generalizes the
notion of a ``spin-network" \cite{15,16} to an arbitrary compact
gauge group and also allows us to give a closed and compact expression
for a spin-network state without referring to a graphical notation 
\cite{15}.

\section{Preliminaries}

We give here only the absolutely necessary information in order to fix
the notation. For further details see \cite{11} and references therein.\\
The space $\agb$ of generalized connections modulo gauge transformations
is the Gel'fand spectrum of the Abelian $C^\star$ algebra generated by the
Wilson-loop functionals for smooth connections, that is, traces of the
holonomy for piecewise analytic loops in the base manifold $\Sigma$. As such
it is a compact Hausdorff space and therefore measures on that space
are in one to one correspondence with positive linear functionals on
$C(\agb)$.\\ 
A certain natural measure $\mu_0$ will play a very crucial role in this 
article so that we go now into more details :\\
In what follows, $\gamma\subset\Sigma$ will always denote a finite, 
unoriented, 
piecewise analytic graph, meaning that it is the union of a finite number 
of analytic edges and vertices. Its 
fundamental group $\pi_1(\gamma)$ is then finitely generated by some loops 
$\beta_1(\gamma),..,\beta_{n(\gamma)}(\gamma)$ which we fix once and for 
all together with some orientation and which are based at some arbitrary but 
fixed basepoint $p\in\Sigma$, $n(\gamma):=\dim(\pi_1(\gamma))$ 
being the number of independent generators of the fundamental group of 
$\gamma$. A function $f$ on $\agb$ is said to be cylindrical
with respect to a graph $\gamma$, $f\in\mbox{Cyl}_\gamma(\agb)$, if it is a 
function only of the 
finite set of arguments $p_\gamma(A):=(h_{\beta_1}(A),..,h_{\beta_n}(A))$
where $h_\alpha(A)$ is the holonomy of $A$ along the loop $\alpha$.
A measure $\mu$ is now specified by its finite joint distributions
$\mu_\gamma$ which are defined by 
\be \label{2.1}
\int_\agb d\mu(A) f(A)=\int_{G^n} d\mu_\gamma(g_1,..,g_n) 
f_\gamma(g_1,..,g_n)
\ee
where $f=f_\gamma\circ p_\gamma$ and $f_\gamma\;:G^n\to\Co$ is a gauge
invariant function. In order 
that this definition makes sense we have to make sure that if we write
$f=f_\gamma\circ p_\gamma=f_{\gamma'}\circ p_{\gamma'}$ in two 
different ways as a cylindrical function where 
$\gamma\subset\gamma'$ is a subgraph of $\gamma'$, then we 
should have that the so-called consistency conditions
\be \label{2.2}
\int_{G^n}d\mu_\gamma f_\gamma=\int_{G^{n'}} d\mu_{\gamma'} f_{\gamma'}
\ee
are satisfied.\\
The natural measure $\mu_0$ is the induced Haar measure, meaning that
$d\mu_{0,\gamma}(g_1,..,g_n)=d\mu_H(g_1)..d\mu_H(g_n)$. One can check that
the consistency conditions are satisfied \cite{2} and that the so defined 
cylindrical measure has a $\sigma$-additive extension $\mu_0$ on the 
projective limit measurable space of the family of measurable spaces 
$\agb_\gamma$ \cite{5}. The space $\agb_\gamma$ 
is defined to be the set of all homomorphisms from the group of based 
loops restricted 
to $\gamma$ into the gauge group modulo conjugation while $\agb$ is the 
set of all homomorphisms
from the whole loop group into $G$ modulo conjugation. Note that the 
semi-group of loops 
with respect to compositions of loops can be given a group structure by 
identifying paths that are traversed in the opposite direction with the
inverse of the original path.

\section{The inverse loop transform}

\subsection{The inverse Fourier transform for compact groups}

Let us recall some basic facts from harmonic analysis on compact gauge groups
\cite{17}. 
\begin{definition}
Let $\{\pi\}$ denote the set of all finite dimensional, non-equivalent
(we fix one representant from each equivalence class once and for all),
unitary, irreducible representations of the compact gauge group $G$, let 
$d_\pi$
be the dimension of $\pi$ and let $\mu_H$ be the normalized Haar measure 
on $G$.\\
For any $f\in L_1(G,d\mu_H)$ define the Fourier transform of $f$ by
\be \label{3.1}
\hat{f}^{ij}_\pi:=\int_G d\mu_H(g) \sqrt{d_\pi}\bar{\pi}_{ij}(g) 
f(g),\;i,j=1,..,d_\pi\; \ee
where $\pi_{ij}(g)$ denotes the matrix elements of $\pi(g)$.
\end{definition}
Note that this definition makes sense because the matrix elements of
$\pi(g)$ are bounded by 1.
\begin{definition}
The Fourier transform of a function is said to be $\ell_1$ or $\ell_2$
respectively iff
\be \label{3.2}
||\hat{f}||_1:=\sum_\pi\sum_{i,j=1}^{d_\pi} \sqrt{d_\pi} 
|f^{ij}_\pi|<\infty\mbox{ or }||\hat{f}||_2:=\sum_\pi\sum_{i,j=1}^{d_\pi} 
|f^{ij}_\pi|^2<\infty  \;. \ee
\end{definition}
The Fourier series associated with a function $f$ on $G$ such that 
$\hat{f}\in\ell_1$ is given by
\be \label{3.3}
\tilde{f}(g):=\sum_\pi\sum_{i,j=1}^{d_\pi} \hat{f}^{ij}_\pi 
\pi_{ij}(g)\sqrt{d_\pi}\;. \ee
The analogue of the Plancherel theorem for $\Rl^n$ is the Peter\&Weyl theorem
\begin{Theorem}[Peter\&Weyl]
1) The functions $g\to\sqrt{d_\pi}\pi_{ij}(g),\;i,j=1,..,d_\pi$ form a 
complete and orthonormal system on $L_2(G,d\mu_H)$.\\
2) For any $f\in L_2(G,d\mu_H)$ it holds that $f=\tilde{f}$ in the sense 
of $L_2$ functions 
and the Fourier transform is a unitary map $\wedge\;:\;L_2(G,d\mu_H)
\to\ell_2$.
\end{Theorem}
The author was unable though to find the analogue of the inverse Fourier
transform for compact groups in the literature which we therefore 
prove here. This theorem answers the question whether a function which is
only $L_1$ can be represented, in the $L_1$ sense, by its Fourier transform.
\begin{Theorem}
Let $f\in L_1(g,d\mu_H)$ such that also $\hat{f}\in\ell_1$. Then 
$f(g)=\tilde{f}(g)$ on $L_1$.
\end{Theorem}
Proof :\\
Let be given any $\hat{k}\in\ell_1$. Then the Fourier series 
\be \label{3.4}
\tilde{f}_k:=\sum_{\pi,i,j}[\sum_k f^{ij}_\pi \bar{k}^{jk}_\pi]\pi_{ij}(g)
\sqrt{d_\pi}\ee 
still converges absolutely as can be seen from the following 
considerations : upon using the Schwarz inequality we obtain the estimate
\[ |\sum_k f^{ik}_\pi \bar{k}^{jk}_\pi|\le\sum_k |f^{ik}_\pi| 
|k^{jk}_\pi|\le\sqrt{\sum_k |f^{ik}_\pi|^2}\sqrt{\sum_l |k^{jk}_\pi|^2}
\le[\sum_k |f^{ik}_\pi|][\sum_l |k^{jl}_\pi|] \]
so that
\be \label{3.5}
||\tilde{f}_k||_1\le\sum_\pi [\sum_{i,k} |f^{ik}_\pi|\sqrt{d_\pi}]
[\sum_{j,l} |k^{jl}_\pi|\sqrt{d_\pi}]\le
[\sum_{\pi,i,k} |f^{ik}_\pi|\sqrt{d_\pi}]
[\sum_{\sigma,j,l} |k^{jl}_\sigma|\sqrt{d_\sigma}]=||\hat{f}||_1 
||\hat{k}||_1 \;.
\ee
Hence (\ref{3.4}) is well-defined and we may write
\ba \label{3.6}
\tilde{f}_k(g) &=&\sum_{\pi,j,k}\sqrt{d_\pi}\bar{k}^{jk}_\pi[\int d\mu_H(h) 
\pi(h^{-1}g)_{kj} f(h)]\nonumber\\
& = & \sum_{\pi,j,k}\bar{k}^{jk}_\pi\sqrt{d_\pi}[\int d\mu_H(h) 
\pi(h^{-1})_{kj}(L_gf)(h)] \nonumber\\
& = & \int d\mu_H(h) (L_gf)(h)\overline{\sum_{\pi,j,k} \sqrt{d_\pi}
k^{jk}_\pi \pi_{jk}(h)}
\nonumber\\
& = & \int d\mu_H(h) (L_gf)(h)\bar{\tilde{k}}(h) \;.
\ea
In the second line we have made use of the translation invariance of the Haar
measure and the unitarity of the representation, $(L_g f)(h)=f(gh)$ is the
definition of the left regular representation of $G$, in the third line
we could switch integration and summation because $\hat{k}\in\ell_1$
and $\bar{\pi}_{jk} L_g f\in L_1(G,d\mu_H)$ 
and finally in the last step we have used the definition of the Fourier
series (\ref{3.3}).\\
We now choose $k^{ij}_\pi:=\sqrt{d_\pi}\pi_{ij}(1)e^{-t\lambda_\pi}$ where
$\lambda_\pi$ are the eigenvalues of the Casimir operator $\Delta$ in the 
representation $\pi$. Then $\tilde{k}(g)=\rho_t(g)$ becomes the heat
kernel on $G$ \cite{20}, that is, the fundamental solution of the equation
\be \label{3.7}
[\frac{\partial}{\partial t}-\Delta]\rho_t(g)=0,\;
\rho_0(g)=\delta_{\mu_H}(g,1) \;. \ee
The motivation for this choice is of course that for $t\to 0$ the lhs
of (\ref{3.6}) tends to $\tilde{f}(g)$ while the rhs should tend to
$f(g)$. Indeed, since $\rho_t$ tends to the $\delta$ distribution on
$G$ wrt the Haar measure, this would be straightforward to see if 
$f\in C^\infty(G)$. To show that this is true even for $f\in L_1(G,d\mu_H)$
we argue as follows : denote $(f\star\rho_t)(g):=\int d\mu_H(h) (L_g 
f)(h)\rho_t(h)$ then we have 
\ba \label{3.8}
||f\star\rho_t-f||_1&=&\int d\mu_H(g)|\int d\mu_H(h) [R_h f-f](g) \rho_t(h)|
\nonumber\\
& \le & \int d\mu_H(g)\int d\mu_H(h) |R_h f-f|(g) \rho_t(h) \ea
where we have used the normalization and positivity of the heat kernel and 
$(R_h f)(g)
=f(gh)$.\\ The idea is now to split the integration domain of the inner
integral into a compact neighbourhood $U$ of the identity of $\mu_H$ volume 
$\delta$ and its complement
$G-U$ in $G$ :
\ba \label{3.9}
||f\star\rho_t-f||_1 &\le &\int_G d\mu_H(g)\int_U d\mu_H(h) |R_h f-f|(g) 
\rho_t(h)\nonumber\\
& & +\int_G d\mu_H(g)\int_{G-U} d\mu_H(h) |R_h f-f|(g) \rho_t(h)
=:I+II \;.
\ea
Since the heat kernel gets concentrated at the identity for
$t\to 0$ we can estimate the second integral while we will estimate the
first integral by a compactness argument.\\
The details are as follows. We first assume that $f\in C(G)$. This 
assumption will be dropped later again. Then f is uniformly continuous  
on the compact set $gU=\{R_hg;\; h\in U\}$ and therefore there is a 
non-negative function $\omega(\delta),\;\lim_{\delta\to 
0}\omega(\delta)=0$ such that $|R_h f-f|(g)\le\omega(\delta)\;\forall
h\in U$. Accordingly 
\be \label{3.10}
I\le\omega(\delta)\int_U d\mu_H(h)\rho_t(h)\le \omega(\delta) \ee
due to the normalization of the heat kernel.\\
The continuous function $(g,h)\to (R_h f-f)(g)\rho_t(h)$ is measurable on 
$G\times G$ wrt the Borel measure $\mu_H\times\mu_H$ (recall \cite{20}
that $\rho_t$ is even real analytic). 
Since $|\int_G d\mu_H(h)\rho_t(h)\int d\mu_H(g)(R_h f-f)(g)|\le 
2||f||_1<\infty$ the theorem of Fubini
allows us to switch the integrations in the second integral and we arrive at
\ba \label{3.11}
II &= &\int_{G-U}d\mu_H(h)\rho_t(h)\int_G d\mu_H(g) |R_h f-f|(g)=
\int_{G-U}d\mu_H(h)\rho_t(h)||R_h f-f||_1\nonumber\\ 
&\le & 
2||f||_1\int_{G-U}d\mu_H(h)\rho_t(h) \;. \ea
Now for any $\epsilon>0$ we find a $\delta(\epsilon)$ such that
$\omega(\delta(\epsilon))<\epsilon/2$ and for this so chosen 
$\delta(\epsilon)$ we find a $t(\epsilon,f)$ so that $2||f||_1\int_{G-U}
d\mu_H \rho_t<\epsilon/2$ since the support of $\rho_t$ gets more and
more concentrated at the identity for $t\to 0$.
Therefore we conclude that for any $f\in C(G),\;\epsilon>0$ there exists a 
$t(\epsilon)>0$ such that $||\rho_t\star f-f||_1<\epsilon$.\\
Now let us focus on a general $f\in L_1(G,d\mu_H)$ and consider an arbitrary
$k\in C(G)$. Then we expand
\be \label{3.11a}
||\rho_t\star f-f||_1\le ||\rho_t\star(f-k)-(f-k)||_1+||\rho_t\star k-k||_1 .
\ee
Now $G$ is a compact Hausdorff space and $\mu_H$ is a Borel measure on 
$G$ so that 
$C(G)$ is dense in $L_1(G,d\mu_H)$ \cite{18}.  We can therefore find a $k$ 
such 
that $||f-k||_1<\epsilon/4$ and therefore $||\rho_t\star(f-k)-(f-k)||\le
2||f-k||_1 ||\rho_t||_1\le\epsilon/2$ for any $t>0$ while we have shown 
above that for any
given $k\in C(G)$ we can always choose $t$ sufficiently small such that 
$||\rho_t\star k-k||_1<\epsilon/2$.\\
This furnishes the proof.\\
$\Box$\\
\\
The theorem can obviously be extended to functions of more than one variable.
This will enable us to define the finite dimensional joint distributions 
of a measure.

\subsection{Computation of the finite joint distributions of a measure}

Recall \cite{26} that every representation of a compact group
is equivalent to a unitary one, so that we may restrict ourselves to 
unitary representations in the sequel. Also,
every such representation is completely reducible. 
In what follows we will assume that we have fixed, in each equivalence class
of irreducible representations that arise in the decomposition into 
irreducibles of a tensor product of irreducible representations (the ones 
that were fixed in definition 3.1), a standard base of independent 
representations which project onto orthogonal representation spaces.
For the case of $SU(2)$ this is the familiar Clebsh-Gordan decomposition
and for $GL(n)$ or $SU(n)$ this can be established, for instance, by choosing
the representations associated with the standard tableaux of the 
corresponding Young diagrammes \cite{26} and for the general case we assume
to have made a similar choice.\\
\\ 
First we introduce a new notion. \\
\begin{definition} 
i) A loop network is a triple $(\gamma,\vec{\pi},\pi)$ consisting of a
graph $\gamma$, a vector $\vec{\pi}=(\pi_1,..,\pi_{n(\gamma)})$ of 
irreducible representations of $G$ and an irreducible representation
$\pi$ of $G$ which takes values in the set of irreducible representations
of $G$ contained in the decomposition into irreducibles of the tensor
product $\otimes_{k=1}^n \pi_k$.\\
ii) A loop-network state is a map from $\agb$ into $\Co$ defined by
\be \label{3.12}
T_{\gamma,\vec{\pi},\pi}(A):=\mbox{\rm tr}[\otimes_{k=1}^{n(\gamma)}
\pi_k(h_{\beta_k(\gamma)}(A))\cdot c(\vec{\pi},\pi)]
\ee
where the matrix $c$ is defined by
\be \label{3.13}
c(\vec{\pi},\pi):=\sqrt{\frac{\prod_{k=1}^{n(\gamma)} d_{\pi_k}}{d_\pi}}
\pi(1)\;. \ee 
\end{definition}
Loop network states satisfy the following important properties.
\begin{Lemma}
i) Given a graph $\gamma$, the set of all loop network states provides 
an orthonormal basis of 
$L_2(\agb_\gamma,d\mu_{0,\gamma})= 
L_2(\agb,d\mu_0)\cap\mbox{Cyl}_\gamma(\agb)$.\\
ii) Given a graph $\gamma'$, consider all its subgraphs $\gamma<\gamma'$.
Remove all the loop network states on $\gamma'$ which are pull-backs 
of loop-network states on $\gamma$. The collection of all loop-network
states so obtained provides an orthonormal basis of $L_2(\agb,d\mu_0)$. 
\end{Lemma}
Proof :\\
i) The orthogonality relations for loop-network states on a given graph 
$\gamma$ follow easily from basic group integration theory. By the 
Peter\&Weyl theorem we have 
\ba \label{3.14}
<T_{\gamma,\vec{\pi},\pi}, T_{\gamma,\vec{\pi}',\pi'}> &=&
\bar{c}(\vec{\pi},\pi)^{(i_1,j_1),..,(i_n,j_n)}
c(\vec{\pi}',\pi')^{(k_1,l_1),..,(k_n,l_n)}\;\times\nonumber\\
&\times& 
\delta_{\vec{\pi},\vec{\pi}'}\frac{1}{\prod_{k=1}^n d_{\pi_k}}
\delta_{i_1,k_1}\delta_{j_1,l_1}\cdot\cdot\delta_{i_n,k_n}\delta_{j_n,l_n}
\nonumber\\
&=& \delta_{\vec{\pi},\vec{\pi}'}\frac{1}{\sqrt{d_\pi d_{\pi'}}}
\mbox{tr}[\pi^\dagger(1)\pi'(1)] \nonumber\\
&=& \delta_{\vec{\pi},\vec{\pi}'} \delta_{\pi,\pi'}
\ea
where we have used that the non-equivalent irreducible -- as well as  
our choice of equivalent -- representations 
of a compact gauge groups are orthogonal, that is, $\pi(1)$ is 
a projector.\\
The completeness of these states on $L_2(\agb_\gamma,d\mu_{0,\gamma})$
follows also from the Peter\&Weyl theorem together with a gauge-invariance
argument :\\
We know that the states 
\be \label{3.15}
T_{\gamma,\vec{\pi}}^{(i_1,j_1)..(i_n,j_n)}:=[\otimes_{k=1}^n 
\pi_k(h_{\beta_k(\gamma)}(A))]^{(i_1,j_1),..(i_n,j_n)}
\ee 
contain an overcomplete set of states for  
$L_2(\agb_\gamma,d\mu_{0,\gamma})$ and thus we only need to
select all the independent gauge invariant combinations of those, that is,
we need to find all the matrices $c$, called contractors, which turn 
(\ref{3.15})
into gauge invariant states when being contracted with them.\\ 
Notice that all the generators $\beta$ are based loops. Therefore under a 
gauge transformation
\be \label{3.16}
\mbox{tr}[T_{\gamma,\vec{\pi}}\cdot c]\to
\mbox{tr}[T_{\gamma,\vec{\pi}}\cdot (\otimes_{k=1}^n\pi_k)(g^{-1})\cdot
c\cdot(\otimes_{k=1}^n\pi_k)(g)]
\ee
and gauge invariance requires choosing $c$ such that
\be \label{3.17}
(\otimes_{k=1}^n\pi_k)(g^{-1})\cdot
c\cdot(\otimes_{k=1}^n\pi_k)(g)=c\mbox{ for all }g\in G.
\ee
Now notice that the matrix $c$ in (\ref{3.16}) is already projected on the
reducible representation space defined by the tensor 
product representation $\otimes_{k=1}^n\pi_k$ because the matrix 
$(\otimes_{k=1}^n\pi_k)(1)$ is a projector on that space
and leaves the matrix $T_{\gamma\vec{\pi}}$ in (\ref{3.15}) invariant under 
multiplication from both sides. It follows that we can expand
\be \label{3.18}
c=\sum_{\pi\in\otimes_{k=1}^n\pi_k} c_\pi
\ee 
where the sum is over the irreducibles contained in the decomposition
of $\otimes_{k=1}^n\pi_k$ into irreducibles and the matrix $c_\pi$
is projected onto the representation space labelled by $\pi$, namely
$\pi(1) c_\pi=c_\pi \pi(1)=c_\pi$. Let us also decompose 
\be \label{3.19}
(\otimes_{k=1}^n\pi_k(g))=\sum_{\pi\in\otimes_{k=1}^n\pi_k} \pi(g)
\ee
We now plug (\ref{3.18}) and (\ref{3.19}) into (\ref{3.17}) and obtain
\be \label{3.20}
\sum_{\pi\in\otimes_{k=1}^n\pi_k} \pi(g)c_\pi\pi(g)^{-1}=
\sum_{\pi\in\otimes_{k=1}^n\pi_k}  c_\pi
\ee
which we multiply by $\pi(1)$ to obtain
\be \label{3.21}
\pi(g)c_\pi\pi(g)^{-1}=c_\pi,
\ee
that is, $c_\pi$ commutes with the representation and therefore must be 
proportional to $\pi(1)$ by the lemma of Schur.\\
\\
ii) It follows immediately from i) that the union of all the loop-network
states for all the graphs $\gamma$ is an overcomplete (uncountable) set
of states on $L_2(\agb,d\mu_0)$ (the graphs label cylindrical functions 
which are dense in $L_2(\agb,d\mu_0)$, compare also \cite{16}). The 
redundant states are eliminated by
the recipe stated in the lemma. In particular then all the representations
involved in $\vec{\pi}$ are required to be non-trivial (except for the 
empty graph) since any loop-network with trivial representations can be 
realized already on a smaller graph. \\
It remains to show that then two 
loop-network states that are defined on different graphs are orthogonal.
But this is trivial because for two graphs $\gamma\not=\gamma'$ there is at
least one generator $\beta$ in which they differ and the representation $\pi$
associated with that generator is non-trivial. Therefore the inner product
between these loop-network states will contain the integral 
$\int d\mu_H(g) \pi(g)=0$. Therefore we get altogether
\be \label{3.22}
<T_{\gamma,\vec{\pi},\pi},T_{\gamma',\vec{\pi}',\pi'}>
=\delta_{\gamma,\gamma'}\delta_{\vec{\pi},\vec{\pi}'}\delta_{\pi,\pi'}\; .
\ee
$\Box$\\
Remark :\\
More concretely, the redundant states can be removed by imposing the 
following constraints on the vector $\vec{\pi}$ : given an edge $e$ of 
$\gamma$ (that is, a maximally analytic piece of $\gamma$) determine
the generators, say $\beta_1,..,\beta_k$, which contain $e$. If $\beta_i$
is cloured with the representation $\pi_i$ then require that the 
tensor product $\pi_1\otimes..\otimes\pi_k$ does not contain any trivial 
representation otherwise the loop-network state would contain a piece 
defined on a smaller graph. \\
This restriction leads to the definition of edge-network states (compare 
the appendix).\\
\\
The next thing to do is to define the Fourier transform of a measure on
$\agb$. 
\begin{definition}
The loop transform (Fourier transform, characteristic functional) of a 
measure $\mu$ on $\agb$ is defined by
\be \label{3.23}
\chi_\mu(\gamma,\vec{\pi},\pi):=<\bar{T}_{\gamma,\vec{\pi},\pi}>:=
\int_\agb d\mu(A) \bar{T}_{\gamma,\vec{\pi},\pi}(A)
\ee
\end{definition}
This definition differs from the one given in \cite{1,9}, however, both 
definitions are equivalent in the sense that they allow for a reconstruction
of $\mu$ according to the Riesz-Markov theorem \cite{18}. Namely, the
former definition is based on the vacuum expectation value of products of
Wilson loop functionals, and according to \cite{1,19}, these functions are
an overcomplete set of functions on $\agb$ (that is, they are subject to
Mandelstam identities) so that we can reexpress them in terms of loop
networks and vice versa.\\
\\
Now let be given a functional $\chi$ on loop-networks. Provided it is 
positive (note that there are no Mandelstam relations between loop
network states any more and that the product of loop network states is a
linear combination of loop network states) we know by the Riesz-Markov
theorem that there is a measure $\mu$ whose Fourier transform is given
by $\chi$. This measure will be known if we know its finite joint 
distributions which automatically form a self-consistent system
of measures whose projective limit (known to exist) gives us back $\mu$.
We now compute these joint distributions.
\begin{Lemma}
If the Fourier transform of a (complex) regular Borel measure $\mu$ on a 
compact gauge group $G$ is in $\ell_1$ then it is absolutely 
continuous with respect to the Haar measure on $G$.
\end{Lemma}
Proof :\\
Given the Fourier coefficients $\chi^{ij}_\pi$ of the measure $\mu$
the Fourier series $\hat{\chi}$ associated with these coefficients  
is an $L_1(G,d\mu_H)$ function due to the anticipated $\ell_1$ property of 
the Fourier coefficients. Moreover, the measure 
$d\hat{\mu}(g):=\hat{\chi}(g)d\mu_H(g)$
has the same Fourier transform as $\mu$. Since the functions defined
by finite linear combinations of the functions $\sqrt{d_\pi}\pi_{ij}$
form a dense set in $C(G)$, G being a compact Hausdorff group,
it follows that both measures define the same
bounded linear functional. Now we infer from the uniqueness part of
the Riesz-Markov theorem that indeed $\mu=\hat{\mu}$ from which
absolute continuity follows. If $\mu$ is even a positive measure then
$\hat{\chi}$ is positive.\\
$\Box$\\
The theorem can obviously extended to any finite number of variables.
\begin{Theorem}
Let $\chi$ be a positive linear functional on $C(\agb)$. Then $\chi$ is
the loop transform of a positive regular Borel measure $\mu$ on $\agb$. If 
for a given graph $\gamma$ with n generators the sequence 
$\{\chi(\gamma,\vec{\pi},\pi)\sqrt{d_\pi\prod_{k=1}^n d_{\pi_k}}\}$ is in 
$\ell_1$ 
then the finite joint distributions of $\mu_\gamma$ are (in 
the sense of $L_1(\agb_\gamma,\mu_{0,\gamma})$) given by 
\be \label{3.24}
\frac{d\mu_\gamma(A)}{d\mu_{0,\gamma}(A)}=
\sum_{\vec{\pi}}\sum_{\pi\in\otimes_{k=1}^n\pi_k} \chi(\gamma,\vec{\pi},\pi)
T_{\gamma,\vec{\pi},\pi}(A) \;.
\ee
\end{Theorem}
Proof :\\
The proof is a straightforward application of the inverse Fourier transform,
theorem 3.2.\\
The convergence condition on the characteristic functional mentioned 
in the theorem together with lemma 3.2 allows us to conclude that
on cylindrical subspaces the measure $\mu_\gamma$ is absolutely continuous 
with respect to
the induced Haar measure $\mu_{0,\gamma}$. 
Therefore there exists a positive $L_1(G^n,d^n\mu_H)$) function 
$\rho_\gamma(g_1,..,g_n)$ such
that it is the Radon-Nikodym derivative of $d\mu_\gamma$ with respect to
$d\mu_{0,\gamma}$ \cite{18}. Since $\mu$ is a gauge-invariant 
measure, the Fourier coefficients of $\rho_\gamma$ satisfy
\be \label{3.24a}
(\otimes_{k=1}^n\pi_k(g^{-1}))_{(i_1,k_1)..(i_n,k_n)}
\rho^{(k_1,l_1)..(k_n,l_n)}_{\gamma,\vec{\pi}}
(\otimes_{k=1}^n\pi_k(g))_{(l_1,j_1)..(l_n,j_n)}
=\rho^{(i_1,j_1)..(i_n,j_n)}_{\gamma,\vec{\pi}}
\ee
so they lie in the invariant subspace spanned by the matrices 
$\pi(1)$ where $\pi\in\otimes_{k=1}^n\pi_k$. Thus
\ba \label{3.25}
\rho^{(i_1,j_1)..(i_n,j_n)}_{\gamma,\vec{\pi}}
& = &\sum_{\pi\in\otimes_{k=1}^n\pi_k} 
\frac{1}{d_\pi}\mbox{tr}[\rho_{\gamma,\vec{\pi}}\pi(1)]
\pi(1)_{(i_1,j_1),..,(i_n,j_n)}(1) \nonumber\\
& = & \sum_{\pi\in\otimes_{k=1}^n\pi_k} 
\frac{1}{\sqrt{d_\pi}}\chi(\gamma,\vec{\pi},\pi)
\pi_{(i_1,j_1),..,(i_n,j_n)}(1) \;.
\ea
Now 
\ba \label{3.26}
& & 
\sum_{i_1,j_1=1}^{d_{\pi_1}}\sqrt{d_{\pi_1}}....\sum_{i_n,j_n=1}^{d_{\pi_n}}
\sqrt{d_{\pi_n}} |\rho_{\gamma,\vec{\pi}}^{(i_1,j_1)..(i_n,j_n)}|
\nonumber\\
& \le &\sum_{\pi\in\otimes_{k=1}^n\pi_k} 
|\chi(\gamma,\vec{\pi},\pi)|
\sum_{i_1,j_1=1}^{d_{\pi_1}}..\sum_{i_n,j_n=1}^{d_{\pi_n}}
|c(\vec{\pi},\pi)_{(i_1,j_1),..,(i_n,j_n)}(1)|\nonumber\\
& = & \sum_{\pi\in\otimes_{k=1}^n\pi_k} 
|\chi(\gamma,\vec{\pi},\pi)| \sqrt{d_\pi\prod_{k=1}^n d_{\pi_k}}
\ea
Therefore the convergence condition on $\chi$ mentioned in the theorem 
implies that $\hat{\rho}_\gamma \in\ell_1$ and the theorem on the inverse 
Fourier transform tells us that in the sense of $L_1$
\ba \label{3.27}
\rho_\gamma(g_1,..,g_n) &=& \sum_{\vec{\pi}}
\sum_{i_1,j_1=1}^{d_{\pi_1}}\sqrt{d_{\pi_1}}..\sum_{i_n,j_n=1}^{d_{\pi_n}}
\sqrt{d_{\pi_n}}\; \times \nonumber\\
& \times & \rho^{(i_1,j_1)..(i_n,j_n)}_{\gamma,\vec{\pi}}
(\otimes_{k=1}^n \pi_k)_{(i_1,j_1)..(i_n,j_n)}(g_1,..,g_n)
\nonumber\\
& = & \sum_{\vec{\pi}} \sum_{\pi\in\otimes_{k=1}^n \pi_k} 
\chi(\gamma,\vec{\pi},\pi) T_{\gamma,\vec{\pi},\pi}
\ea
where we used (\ref{3.25}) and the definition of a loop-network. This 
furnishes the proof.\\
$\Box$\\
Note that if we knew that $\rho_\gamma\in 
L_2(\agb_\gamma,d\mu_{0,\gamma})$ then we could have simply made use 
of the fact that loop networks provide for an orthonormal basis of 
$L_2(\agb,d\mu_0)$ to conclude theorem 3.3 directly from the gauge
invariant version of the Peter\&Weyl theorem. This is, however, not
necessarily the case.

\section{Examples of inverse Fourier transforms}

In order to determine whether a given function $\chi$ from (singular) knots 
into the complex numbers 
arises as the loop transform of a measure one has to check two things :\\
1) All the identities that are satisfied by products of traces of holonomies
of loops have to be satisfied Mandelstam identities \cite{19}. 
Alternatively, it has to be true that $\chi$ can be written purely in 
terms of loop-network states.\\
2) It is a positive linear functional on any cylindrical subspace of
$C(\agb)$.\\
a) An example of a (singular) knot function that satisfies these criteria is 
of course the Fourier 
transform of any $\sigma$-additive measure on $\agb$. Let us look 
at the Fourier transform of the measure $\mu_0$ which is even diffeomorphism
invariant so that $\chi$ is a singular knot invariant :
\be \label{4.1}
\chi_{\mu_0}(\gamma,\vec{\pi},\pi)=\delta_{\vec{\pi},\vec{0}}\delta_{\pi,0}
\ee
where $0$ denotes the trivial representation. In other words, 
$\chi$ is non-vanishing only on the trivial loop network $1$. Therefore we 
find for the
finite joint distribution precisely $\rho_\gamma(g_1,..,g_n)=1$ according to
(\ref{3.24}).\\
b) A second example of a singular knot invariant is given as follows :\\
Let $K$ be a regular knot invariant and let orb$(\alpha_0)$ be the orbit
of the regular knot $\alpha_0$ under the diffeomorphism group under 
question. Then the formal expression \be \label{4.2}
d\mu(A):=\sum_{\alpha\in\mbox{orb}(\alpha_0)} \bar{T}_\alpha(A) d\mu_0(A)
\ee
is a diffeomorphism invariant measure on $\agb$ where $T_\alpha=\mbox{tr}
h_\alpha$, namely its Fourier transform
\be \label{4.3}
\chi_\mu(\gamma,\vec{\pi},\pi)=\chi_{\mbox{orb}(\alpha_0)}(\gamma)
\delta_{\vec{\pi},def}\delta_{\pi,def}
\ee  
($\chi_S$ means the characteristic function of a set $S$)
is a  bounded linear functional on $C(\agb)$ and thus by the extension
of the Riesz-Markov theorem \cite{18} we know that it corresponds to
a unique complex regular Borel measure on $\agb$ which is rigorously 
defined. So we get a new singular knot invariant from a regular one !\\
c) Now we provide an example of a Fourier transform for a 
non-diffeomorphism invariant measure :\\
In two Euclidean spacetime dimensions one can choose the generators 
of a graph to be simple, meaning that they do not have 
self-intersections, and non-overlapping, meaning that the intersection of 
the 
surfaces that any two of them enclose have zero Euclidean area \cite{10}.
Then the characteristic functional (in the continuum) for pure Yang-Mills
theory on the Euclidean plane is given by 
\be \label{4.4}
\chi_{\mu_{YM}^{(2)}}(\gamma,\vec{\pi},\pi)=e^{-\frac{g_0^2}{2}\sum_{k=1}^n
\lambda_{\pi_k}\mbox{Ar}(\beta_k)}\sqrt{d_\pi\prod_{k=1}^n d_{\pi_k}}
\ee
where $g_0$ is the bare coupling constant, $-\lambda_\sigma$ is the 
eigenvalue of the Casimir operator on the matrix element functions  
$\sigma_{ij}(g)$ for an irreducible representation $\sigma$ and Ar$(\alpha)$
is the area of the surface enclosed by a simple loop $\alpha$.\\
Let us compute, for instance, the one-dimensional joint distributions.
We find from (\ref{3.24}) (if $\vec{\pi}$ is one dimensional then of course
also $\pi=\vec{\pi}$ is the only possible choice) 
\be \label{4.5}
\rho_{\gamma=\beta}(g)=\sum_\pi d_\pi e^{-\frac{g_0^2}{2}\mbox{Ar}(\beta)
\lambda_\pi}\chi_\pi(g)
\ee
therefore $\rho_{\gamma=\beta}(g)=\rho_{g_0^2\mbox{Ar}(\beta)}(g)$ where,
as before, $\rho_t$ is the heat kernel on $G$.\\
\\
\\
\\
{\large Acknowledgements}\\
\\
This research project was supported by NSF grant PHY93-96246,
the Eberly research fund of the Pennsylvania State University and
grant DE-FG02-94ER25228 of Harvard University.

\begin{appendix}

\section{Edge networks}

In this appendix we show how the developments of this paper can be 
written in terms of edges which are more convenient to deal with if\\
a) one wants to write down a complete orthonormal basis \cite{15,16} of
$L_2(\agb,d\mu_0)$ without having to make use of the recipe 
mentioned in lemma 3.1,ii) and\\
b) if one is interested in applications to quantum gravity, in
particular if one is to obtain the spectrum of certain area and
volume operators \cite{11,24}.\\
\\
Again we consider an unoriented graph $\gamma$ and fix an 
orientation  
for each of its edges once and for all. 
Note that we do not have to make a choice of edges in this case 
because they are defined to be the maximally analytic pieces of the
given graph. Denote by $E_\gamma$ and $V_\gamma$ the set of
edges and vertices of $\gamma$ respectively, $n_E(\gamma)$ and
$n_V(\gamma)$ are the number of these edges and vertices respectively  and in 
general we 
will denote edges by the symbol $e$ and vertices by the symbol $v$.
\begin{definition} 
i) An edge network is a triple $(\gamma,\vec{\pi},\vec{\sigma})$ 
consisting of a graph $\gamma$ an edge vector of irreducible 
non-trivial representations
$\vec{\pi}=(\pi_1,..,\pi_{n_E(\gamma)})$ and a vertex vector of
irreducible trivial representations 
$\vec{\sigma}=(\sigma_1,..,\sigma_{n_V(\gamma)})$.
The irreducible representation $\sigma_v$ takes values in the set of
trivial irreducible representations that are contained in the decomposition 
into 
irreducibles of $\otimes_{e^-=v} \bar{\pi}_e\otimes_{e^+=v} \pi_e,\;
e^\pm$ being the starting or ending point of $e$ (these representations
are automatically orthogonal to each other). We assume that $\vec{\pi}$
is such that the space of possible $\vec{\sigma}$ is non-empty.\\
ii) An edge network state is a function from $\agb$ into the complex numbers
defined by
\be \label{a.1}
T_{\gamma,\vec{\pi},\vec{\sigma}}(A):=\mbox{tr}[\otimes_{e\in E_\gamma}
\pi_e(h_e(A)) \cdot c(\gamma,\vec{\pi},\vec{\sigma})]
\ee
where the matrix $c$ is defined as follows :\\
There exist permutation matrices $P_\gamma^\pm$ such that
\be \label{a.2}
(P_\gamma^\pm)^{-1}\cdot\otimes_{e\in E_\gamma} 
\pi_e(h_e)\cdot P_\gamma^\pm=
\otimes_{v\in V_\gamma}\otimes_{e^\pm=v}\pi_e(h_e) 
\ee
then 
\be \label{a.3}
c(\gamma,\vec{\pi},\vec{\sigma}):=\sqrt{\prod_{e\in 
E_\gamma} d_{\pi_e}} P_\gamma^-\cdot \otimes_{v\in V_\gamma} 
c_v(\vec{\sigma})\cdot (P_\gamma^+)^{-1}
\ee
where the vertex contractor is given by
\be \label{a.4}
c_v(\vec{\sigma})_{ij}:=\frac{\sigma_v(1)_{(i_0j_0),(ij)}}
{\sqrt{\sigma_v(1)_{(i_0j_0),(i_0 j_0)}}}
\ee
and where $(i_0 j_0)$ is an arbitrary but fixed choice of index pairs 
such that the denominator of (\ref{a.4}) is
non-vanishing (since 
$\sigma_v$ is a trivial representation, any choice of $i_0,j_0$ leads to
the same vector $c_v$ up to a multiple constant). Here 
the index structure comes from $[\otimes_{e^-=v}\bar{\pi}_e]\otimes
[\otimes_{e^+=v} \pi_e]_{(ij),(i_0 j_0)}$. 
\end{definition}
The composition of loops can result in a loop that is defined already on 
a smaller graph. This is the source of the redundancy mentioned in 
lemma 3.1. Something similar cannot happen with edges whence there is
no redundancy in the definition of edge networks.\\
In the special case of $G=SU(2)$ the 
notion of edge-networks coincides with the notion of spin-networks
\cite{15,16}. In this case the vertex contractors can easily be seen to be
the usual Clebsh-Gordan coefficients $c_v=<0,\{(j_v,l_v)\}>$
for the addition of all the angular momenta corresponding to the irreducible 
representations with which the edges starting and ending at $v$ are
coloured. Note that our analytical expression (\ref{a.1}) for an 
edge-network state
does not need any graphical visualization and no preferred role is played by
trivalent graphs \cite{15}. 
\begin{Theorem}
The set of all edge networks provides for a complete orthonormal basis
of $L_2(\agb,d\mu_0)$.
\end{Theorem}
Proof :\\
1) Orthonormality :\\
Note that $P_\gamma^T=P_\gamma^\dagger=P_\gamma^{-1}$ for any permutation 
matrix $P_\gamma$ which merely reshuffles the order of the factors in 
the tensor product $\otimes_{e\in E_\gamma} \pi_e$. Therefore
\ba \label{a.5} 
<T_{\gamma,\vec{\pi},\vec{\sigma}},T_{\gamma',\vec{\pi}',\vec{\sigma}'}>
&=& \delta_{\gamma,\gamma'}\delta_{\vec{\pi},\vec{\pi}'}
\mbox{tr}[c(\gamma,\vec{\pi},\vec{\sigma})^\dagger 
c(\gamma,\vec{\pi},\vec{\sigma})] \nonumber\\
&=&\delta_{\gamma,\gamma'}\delta_{\vec{\pi},\vec{\pi}'}
\mbox{tr}[(P_\gamma^-\cdot \otimes_{v\in V_\gamma} 
c_v(\vec{\sigma})\cdot (P_\gamma^+)^{-1})^\dagger \cdot\nonumber\\
&& P_\gamma^-\cdot \otimes_{v\in V_\gamma} 
c_v(\vec{\sigma}')\cdot (P_\gamma^+)^{-1})]\nonumber\\
&=&\delta_{\gamma,\gamma'}\delta_{\vec{\pi},\vec{\pi}'}
\mbox{tr}[\otimes_{v\in V_\gamma} 
(c_v(\vec{\sigma}))^\dagger c_v(\vec{\sigma}')]\nonumber\\
&=&\delta_{\gamma,\gamma'}\delta_{\vec{\pi},\vec{\pi}'}
\delta_{\vec{\sigma},\vec{\sigma}'}
\ea
we have orthonormality. We have used that two different graphs differ
in at least one edge which carries a non-trivial irreducible representation
and therefore the integral (\ref{a.5}) with respect to the Haar measure 
vanishes as well as the reality and symmetry of the projectors
$\pi(1)$ for any irreducible representation $\pi$.\\
2) Completeness :\\
As in (\ref{3.15}) we start from the observation that the states 
\be \label{a.6}
(T_{\gamma,\vec{\pi}})_{(i_1 j_1)..(i_{n(E_\gamma)} j_{n(E_\gamma)})}
=(\otimes_{e\in E_\gamma} 
\pi_e(h_e))_{(i_1 j_1)..(i_{n(E_\gamma)} j_{n(E_\gamma)})}
\ee
contain a complete set of states for $L_2(\agb_\gamma,d\mu_{0,\gamma})$
so that we only need to contract (\ref{a.6}) in all the possible gauge 
invariant ways. Making use of (\ref{a.2}) we see that under a gauge 
transformation $g_v$ at every vertex $v$ we have 
\ba \label{a.7}
\mbox{tr}[T_{\gamma,\vec{\pi}}\cdot c] &\to&
\mbox{tr}[T_{\gamma,\vec{\pi}}\cdot 
P_\gamma^-\cdot [\otimes_v\otimes_{e^-=v}
\pi_e(g_v^{-1})]\cdot \nonumber\\
& & [(P_\gamma^-)^{-1}\cdot c \cdot 
P_\gamma^+]\cdot [\otimes_v\otimes_{e^+=v}
\pi_e(g_v)]\cdot (P_\gamma^+)^{-1}]
\ea
so that gauge invariance requires that
\be \label{a.8}
[\otimes_v\otimes_{e^-=v}\pi_e(g_v^{-1})]\cdot 
[(P_\gamma^-)^{-1}\cdot c \cdot   P_\gamma^+]
[\otimes_v\otimes_{e^+=v}\pi_e(g_v)]
=[(P_\gamma^-)^{-1}\cdot c\cdot P_\gamma^+]\;.
\ee
Without loss of generality we write $c=P_\gamma^-\cdot\sum_{i=1}^n
\otimes_v s_v^{(i)} \cdot (P_\gamma^+)^{-1}$ for some $n$ 
and arrive at 
\be \label{a.9}
\sum_i \otimes_v s_v^{(i)}=\sum_i 
\otimes_v [\otimes_{e^-=v}\pi_e(g_v^{-1})]\cdot 
s_v^{(i)} \cdot [\otimes_{e^+=v}\pi_e(g_v)]
\ee 
which we can write more conveniently as the eigenvalue equation
\be \label{a.10}
\sum_i \otimes_v s_v^{(i)}=[\sum_i\otimes_v s_v^{(i)}]\cdot 
[\otimes_v \nu_v(g_v)]
\ee
where we have abbreviated
\be \label{a.11}
\nu_v(g_v)=[\otimes_{e^-=v}\bar{\pi}_e \otimes_{e^+=v}\pi_e](g_v).
\ee
Since $\nu_v(1)$ is a projector, it follows that the eigenvector
on the lhs of (\ref{a.10}) lies in the (reducible) subspace
corresponding to $\otimes_v \nu_v$, that is, each $s_v^{(i)}$ lies in the 
(reducible) subspace corresponding to $\nu_v$.\\
We now ask for the possible solutions $x$ of the equation
$x[\otimes_v \nu_v(g_v)]=x\;\forall (g_v)_{v\in V_\gamma}\in 
G^{n(V_\gamma)}$. Setting $g_{v'}=1$ for all $v'\in V_\gamma$ except for 
one, $v$ say, shows that $x$ has to project onto one of the trivial 
representations $\sigma_v$ contained in $\nu_v$. Therefore $x$ has the 
general form
$\sum_{\vec{\sigma}\in\vec{\nu}=trivial} k_{\vec{\sigma}}
\otimes_v c_v(\vec{\sigma})$
for some complex numbers $k_{\vec{\sigma}}$ which is 
precisely of the form (\ref{a.4}). 
Since the space of vectors spanned by the matrix elements of the trivial
representation is one-dimensional the choice made in (\ref{a.4})
means no loss of generality.
This proves completeness.\\
$\Box$\\
It is straightforward to see how the inverse loop transform theorem
can be translated into an ``inverse edge transform" theorem.

\end{appendix}

\end{document}